\documentclass[a4paper]{article}

\usepackage{icrc2013}
\usepackage[T1]{fontenc}
\usepackage[utf8]{inputenc}
\usepackage{lmodern}
\usepackage[francais, english]{babel}

\title{Improved sensitivity of H.E.S.S.-II through the fifth telescope focus system}

\shorttitle{}

\authors{
F. Krayzel$^{1}$,
G. Maurin$^{1}$,
L. Brunetti$^{1}$, 
J.-M. Dubois$^{1}$, 
A. Fiasson$^{1}$,
L. Journet$^{1}$, 
G. Lamanna$^{1}$, 
T. Leflour$^{1}$,  
B. Lieunard$^{1}$,
I. Monteiro$^{1}$,
S. Rosier-Lees$^{1}$, 
for the H.E.S.S. Collaboration.
}

\afiliations{
$^1$ LAPP, Universit\'{e} de Savoie, CNRS/IN2P3, F-74941 Annecy-le-Vieux, France
. \\
}

\email{fabien.krayzel@lapp.in2p3.fr}

\abstract{The Imaging Atmospheric Cherenkov Telescope (IACT) works by imaging the very short flash of Cherenkov radiation generated by the cascade of relativistic charged particles produced when a TeV gamma ray strikes the atmosphere. This energetic air shower is initiated at an altitude of 10-30 km depending on the energy and the arrival direction of the primary gamma ray. Whether the best image of the shower is obtained by focusing the telescope at infinity and measuring the Cherenkov photon angles or focusing on the central region of the shower is a not obvious question. This is particularly true for large size IACT for which the depth of the field is much smaller. We address this issue in particular with the fifth telescope (CT5) of the High Energy Stereoscopic System (H.E.S.S.); a 28 m dish large size telescope recently entered in operation and sensitive to an energy threshold of tens of GeVs. CT5 is equiped with a focus system, its working principle and the expected effect of focusing depth on the telescope sensitivity at low energies (50-200 GeV) is discussed.

}


\begin{document}
\maketitle

\section{Introduction}

IACTs detect gamma rays indirectly: when a gamma ray reaches the atmosphere, the latter behaves as a calorimeter and an air shower starts. The Cherenkov light is emitted by the secondary charged particles from the shower. This light occurs in the ultraviolet and visible wavelengths range. Then, an image of the light pool is obtained thanks to the mirrors which collect the photons and focus them to the camera, which is typically an array of hundreds of photomultipliers. Since the telescope mirrors observe the Cherenkov light at a finite distance, it is relevant to focus on the atmosphere rather than to infinity.
  
Moreover, the maximum Cherenkov light yield occurs when the air shower transversal extension has reached its maximum. The height of such a maximum depends on the energy of the gamma ray and increases with the zenith angle arrival direction.
If the camera is focused to infinity, it is expected to collect more shower photons spread over a larger light pool, broadening the image and making worsen the cosmic-ray rejection capability and the direction reconstruction of the instrument.
In this work the camera focus system instrumenting the H.E.S.S.-II CT5-28 m telescope is described. We report the results of a Monte Carlo simulation study investigating the effect of the focus position of its camera on the telescope sensitivity.

\section{The H.E.S.S.-CT5 focus system}

\begin{figure}[h]
  \centering
  \includegraphics[width=0.32\textwidth, angle = 270]{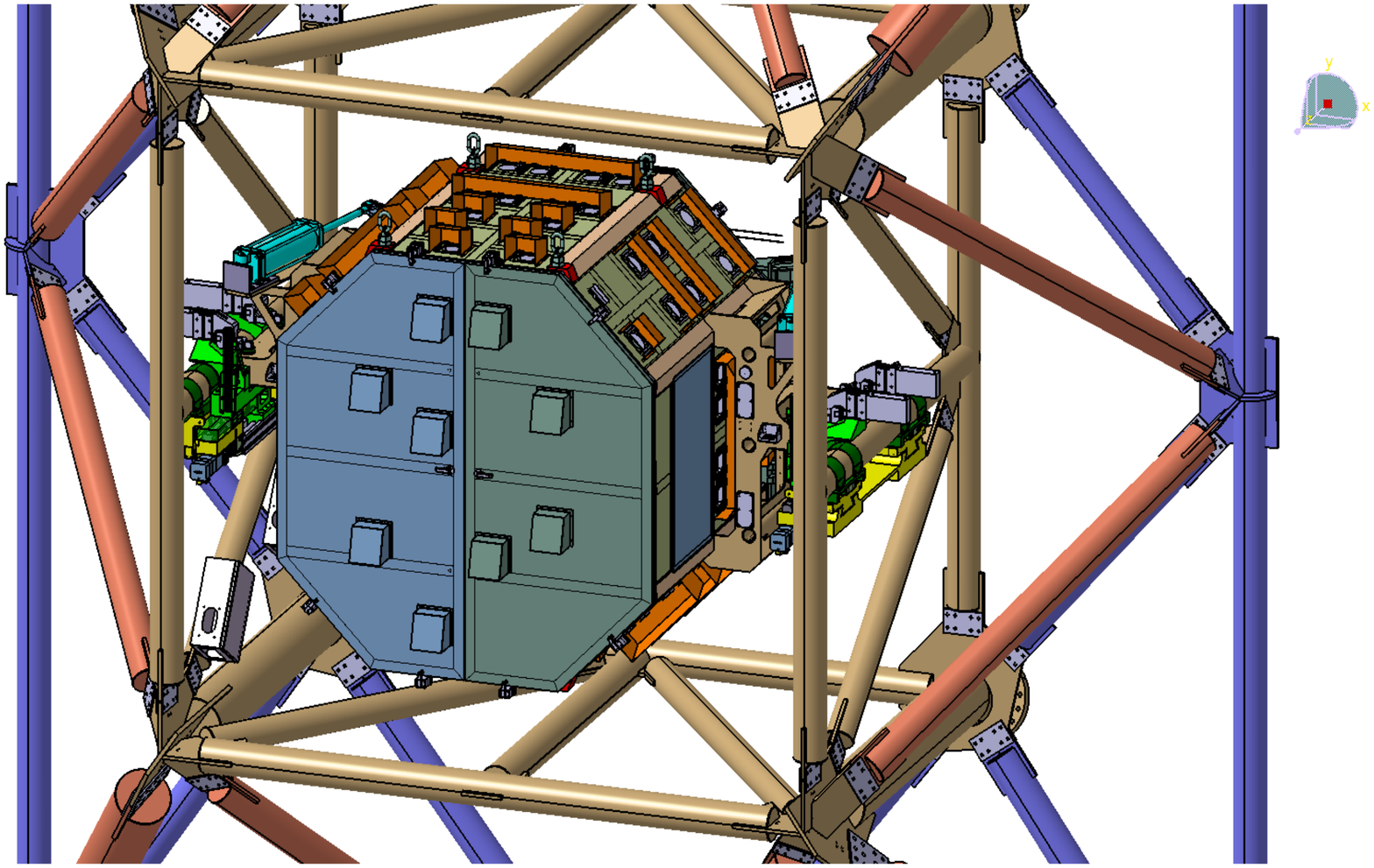}
  \caption{CAD (Computer-Aided Design) images of the CT5 camera loaded onto the support mast structures of the telescope and locked by the focus system.}
  \label{large_CAD}
 \end{figure}

\begin{figure*}[!t]
  \centering
  \includegraphics[width=\textwidth, angle = 180]{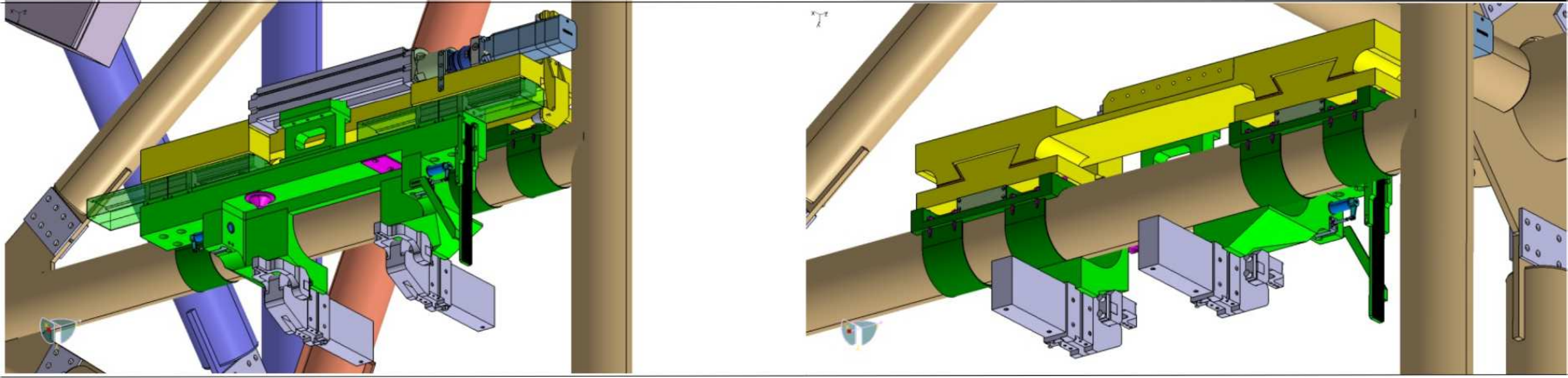}
  \caption{CAD images: in yellow the fixed part and in green the moving part.}
  \label{small_CAD}
 \end{figure*} 
 
\begin{figure}[!ht]
  \centering
  \includegraphics[width=0.45\textwidth, angle = 0]{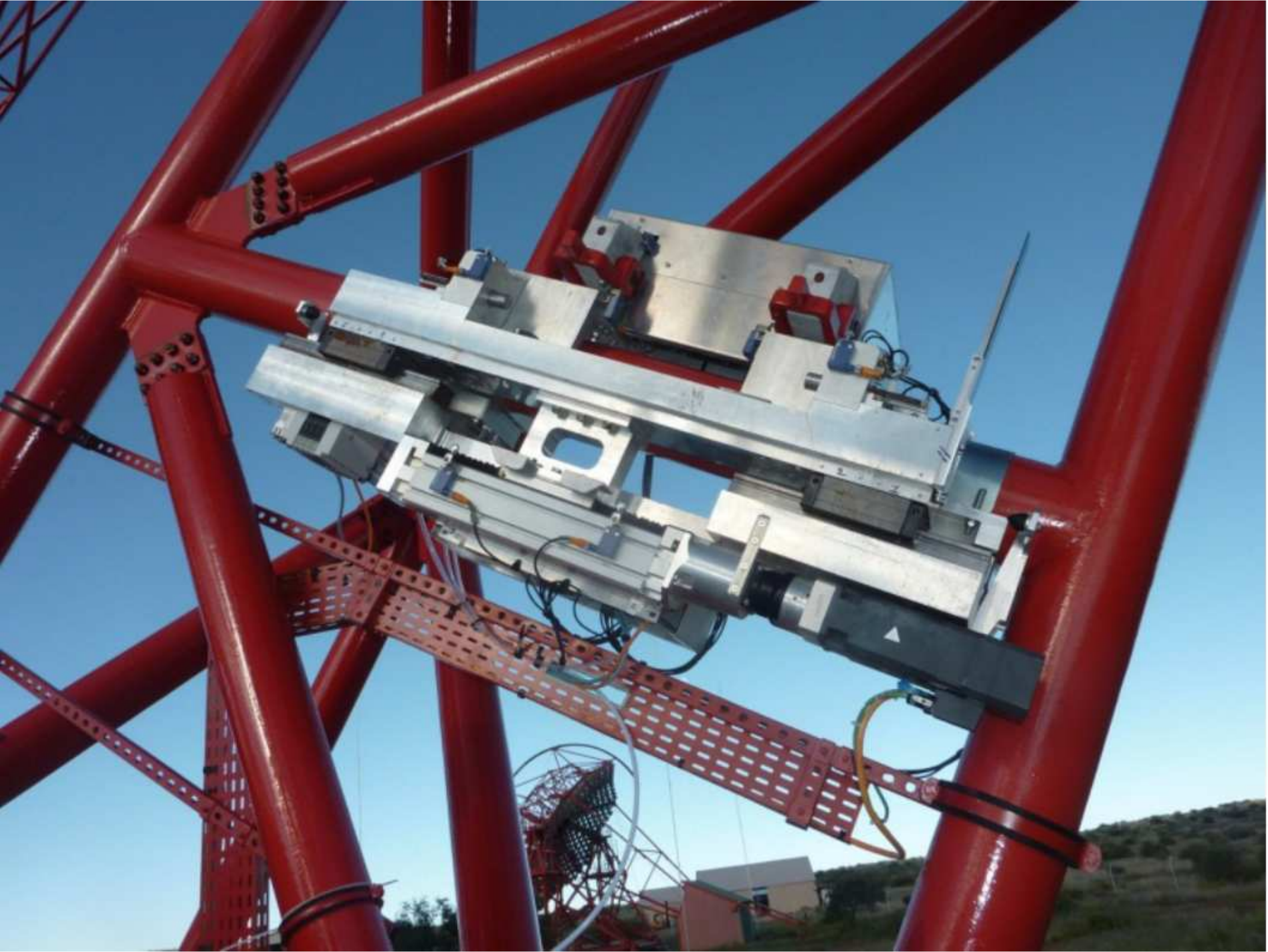}
  \caption{Picture of the CT5 focus system.}
  \label{pic}
 \end{figure}

H.E.S.S.-II, the second phase of the H.E.S.S. experiment, is an array of five Very High Energy (VHE) gamma-ray Cherenkov telescopes (four having 13 m diameter mirror dish and one, namely CT5, with a 28 m dish) situated in the Khomas Highland of Namibia. In particular the fifth telescope, CT5, entered in operation in 2012 and enables to lower down the minimal energy threshold from a few hundreds of GeV to $\sim$30 GeV. The nominal focal distance between the position of the camera and the centre of the mirror dish is 36 m. Unlike the remaining four smaller H.E.S.S. telescopes, the focal distance of CT5 is not fixed but adjustable thanks to a slewing system of the camera along the focal axes of the telescope. The focus system has been designed to operate at 1800 m altitude and to respect a series of critical environmental requirements related to wind, rain and dust. Its main technical specifications are:
\begin{itemize}
\item maximum camera shift range of 230 mm;
\item focal distance range from 35.93 m to 36.16 m; 
\item displacement accuracy of 1 mm;
\item 3000 kg total weight of the focal equipment (namely the camera and the focus system) to be shifted; 
\item full working efficiency guaranteed for varying telescope pointing positions from -35$^{\circ}$ to +180$^{\circ}$ (with respect to the horizontal axis);
\item operating within a temperature range from 0 $^{\circ}$C to 40 $^{\circ}$C.
\end{itemize}
The mechanical design of the focus system aims at two main functionalities: 
\begin{enumerate}
\item the camera installation and locking into the masts structures with an accuracy of 0.35 mm. For maintenance or calibration purposes as well as in case of not favourable weather conditions, the camera is used to be loaded and unloaded from the CT5 telescope by means of a dedicated mecatronic and semi-automatic system (which is not described here). The camera is used to be locked and unlocked to the masts by using a pneumatic system composed of four toggle fasteners and four jacks. For safety reasons, the control of this system is hardware conditioned and each toggle fastener is able to manage 1.5 t whereas the whole mass of the camera and the focus system together is equal to 3.5 t. Figure \ref{pic} shows the final focus system on the telescope.

\item adjusting the position of the camera (the dimensions of which are 2.3 m $\times$ 2.3 m $\times$ 2.5 m) along the optical axis, inside the camera support mast structures (a square cuboid structure of 3.9 m $\times$ 3.9 m $\times$ 2.5 m made of steel pipes and shown in Figure \ref{small_CAD}). The focus system is composed of two parts, one fixed and locked on the telescope thanks to four steel rings and a moving one in order to ensure the translation of the camera (see Figure \ref{large_CAD}). These two parts are linked by four rail-base devices. The range is 200 mm and is carried out by two ball-screws (of 5 mm pitch). The displacement is managed by two independent brushless motors equipped by absolute multiturn encoders which guarantee a speed motion of 0.015 m s$^{-1}$ and an accuracy of 0.05 mm.
\end{enumerate}


The two homologue focus system moving components, laying along two opposite lateral sides of the camera are completely automatized. They are both in charge of executing the translation of the camera along two parallel axes but not mechanically coupled. The control set point and synchronization of both axes is done via a virtual central axis which is the master of both displacements, while the two mechanical axial translators are slaves of it. The focusing process is required to be guaranteed until the end of the programmed translation even in case of unexpected blocking cause as power cut. For such a purpose the system is controlled through a Programmable Logic Controller architecture composed of two Central Processing Units
(one in charge of the classical locking and unlocking cycles of the camera and one more in charge of the management of the camera translations), two field-buses dedicated  respectively to the deported devices (input output modules, variators and control touch panels) and to specific variators in charge of managing  power cuts and synchronized displacements.

\begin{figure}[!ht]
  \centering
  \includegraphics[width=0.21\textwidth, angle = 0]{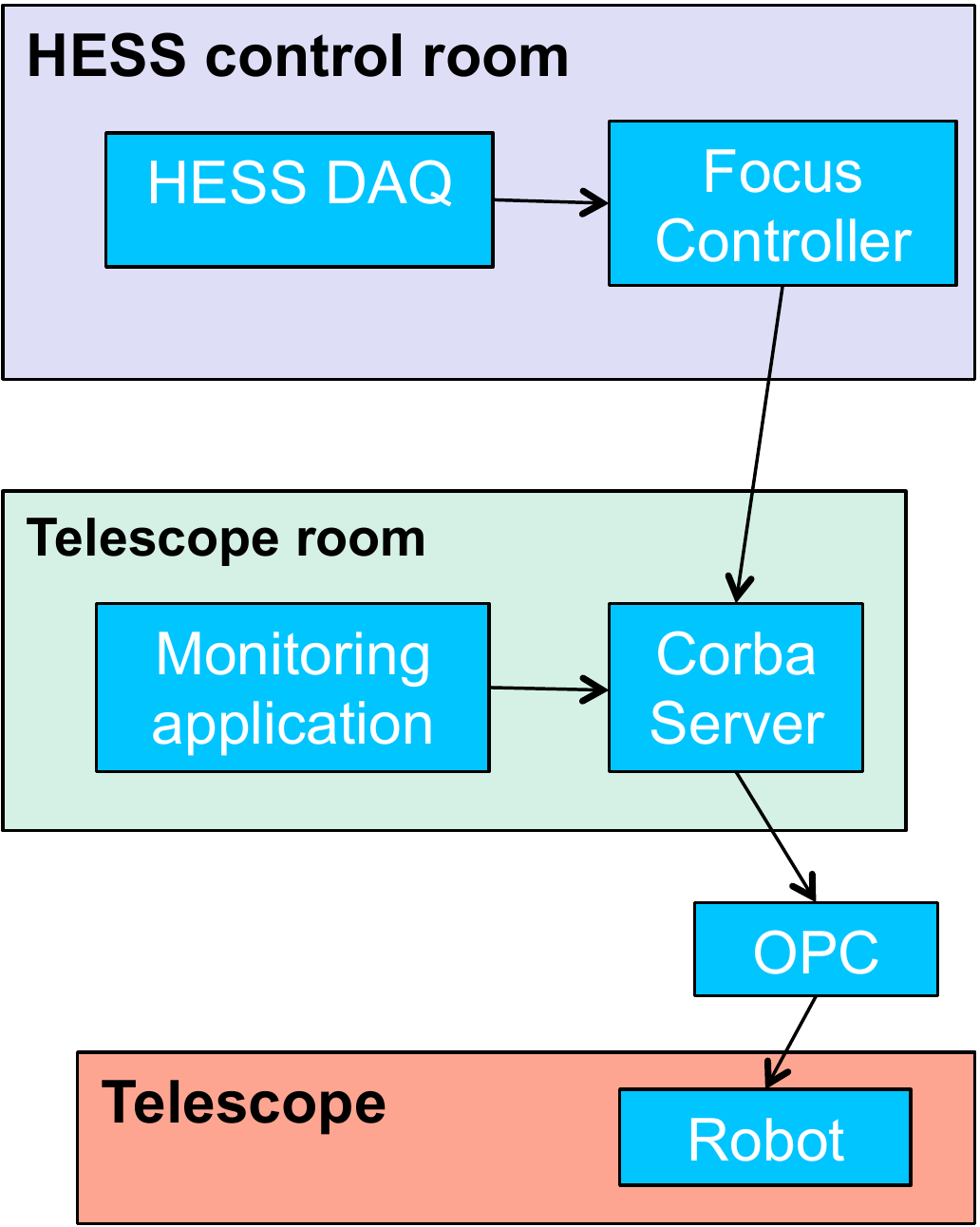}
  \caption{Sketch of the control system.}
  \label{schema}
 \end{figure}

A dedicated controller and a Corba server allow the data acquisition software to control the camera position via a OLE for Process Control (OPC) as represented on Figure \ref{schema}. Moreover, a Java application monitors and controls this system with different levels of expertise.

\begin{figure*}[!t]
  \centering
  \includegraphics[width=0.3\textwidth, angle = 270]{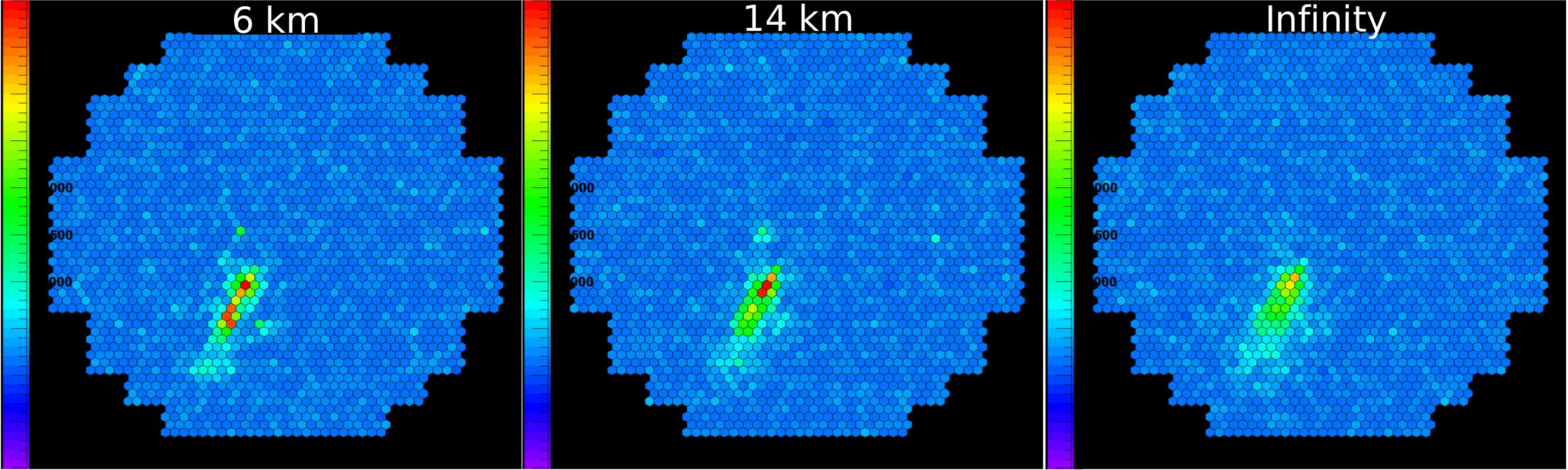}
  \caption{Event display of a simulated gamma ray at 200 GeV.}
  \label{ED}
 \end{figure*}

\section{Simulations}

Some Monte Carlo simulation studies were conducted with the purpose of investigating and validating the working principle of the focus system and its effect on the CT5 telescope sensitivity. 

These studies have been done using the official H.E.S.S. simulation software packages. First of all a total amount of $\sim$1.4 million primary gamma rays were generated and shared among several fixed energy samples (between 12 GeV and 5 TeV) and for three different incident zenith angles (ZA): 0$^{\circ}$, 32$^{\circ}$ and 60$^{\circ}$. The gamma-rays induced air showers were simulated with the KASKADE software package \cite{bib:kaskade} which produces and propagates the final Cherenkov photons on ground as well. Then the telescope simulator reproduces the Cherenkov light reflection through the telescope optics and the camera response (i.e. photodetectors and associated front-end electronics). The shadow of the telescope structure (mast and camera) the effective reflectivity of the mirrors as a function of the photon wavelength are well taken into account. A varying focus height between 5 km and 26 km, plus one more focus condition corresponding to infinity were simulated. Then photons are propagated to the 
pixels of the camera considering the focus position. The time delay between the mirrors and the camera is computed for each photon. At this stage the simulated electronic acts.

Finally whenever the trigger conditions are fulfilled an equivalent \emph{raw} event-base camera image is registered and the event reconstruction methods in use in H.E.S.S. are applied to it.

\section{Results}

\begin{figure}[h]
  \centering
  \includegraphics[width=0.5\textwidth, angle = 0]{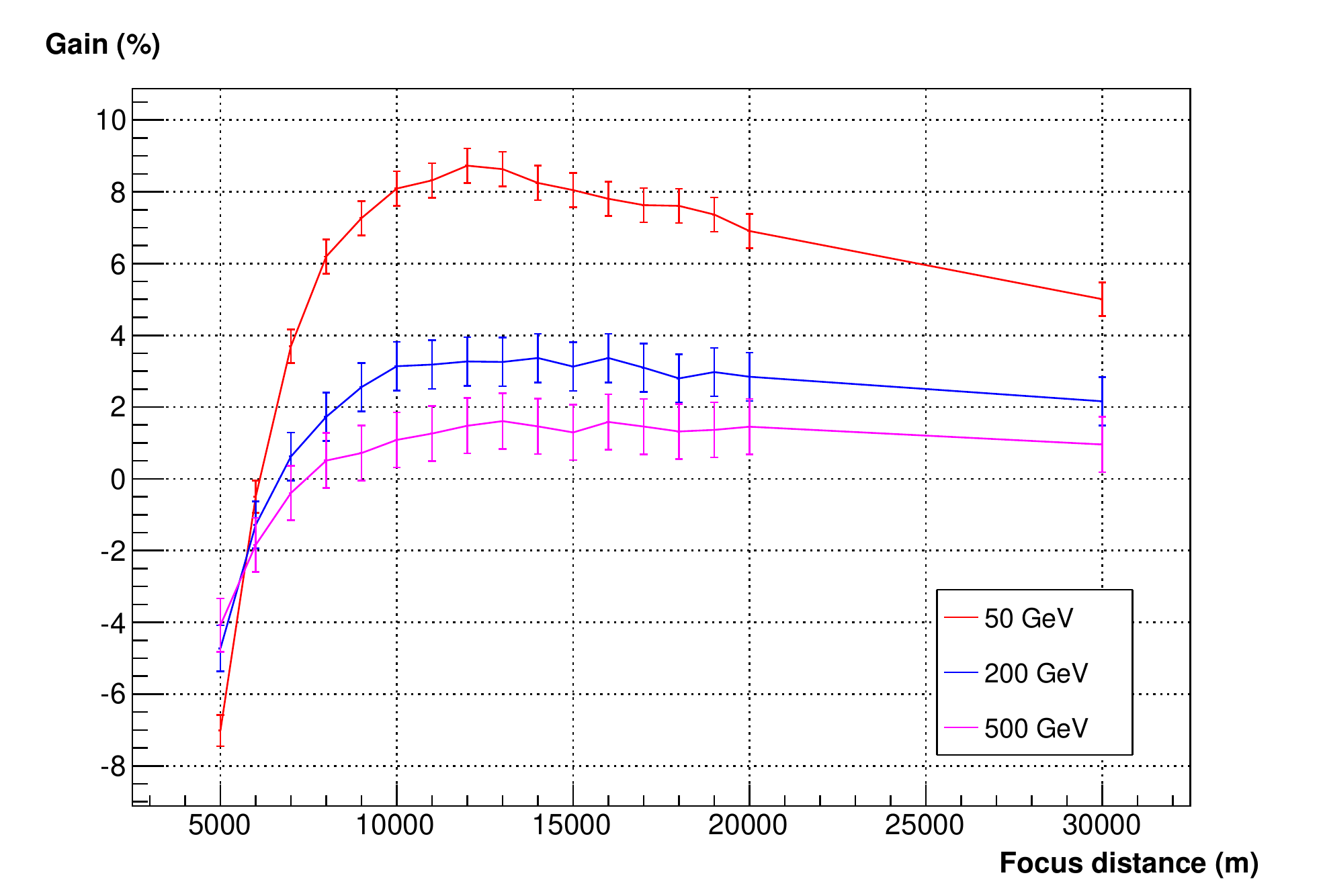}
  \caption{Trigger rate gain in percentage as a function of focus position and relative to the case of focusing to infinity. Gamma-ray samples of three different energies and for incident direction with zenith angle = 0$^{\circ}$.}
  \label{gain0}
 \end{figure}

In order to estimate the effect of the focus, we first limit our analysis to the trigger rates without considering the effect of the event reconstruction. Figure \ref{ED} shows a typical camera shower image from a primary gamma ray of 200 GeV. The three pictures represent the camera pixels for respectively three different focus positions: 6 km, 14 km and infinity. As expected, we observe a spread when the focus is at infinity.
The trigger effective area has been deduced after simulations for each focusing position and for different gamma-ray incident angles. The results are proposed in figure \ref{gain0} relative to the values obtained in the case of focus at the infinity and zenith angle equal zero.
We choose four fixed energy gamma-ray samples (50 GeV, 200 GeV, 500 GeV and 1250 GeV) and vary the focus from 5 km to 20 km. The results are the following :
\begin{itemize}
\item the gain is more important at low energy. These energies are near the energy threshold of the telescope and a spread of the light pool imply to go below this limit.
\item there is a gain for all energies. As expected the spread of the light pool is a detrimental effect for each energy. 
\item the performances are very stable in a large focus position range. This is because the spread of the light remains within the pixel size.
\item at very low height there is a reduction of the performances for all energies. It means that the focus position is far below the shower maximum for each chosen energy.
\end{itemize}

\begin{figure}[h]
  \centering
  \includegraphics[width=0.5\textwidth, angle = 0]{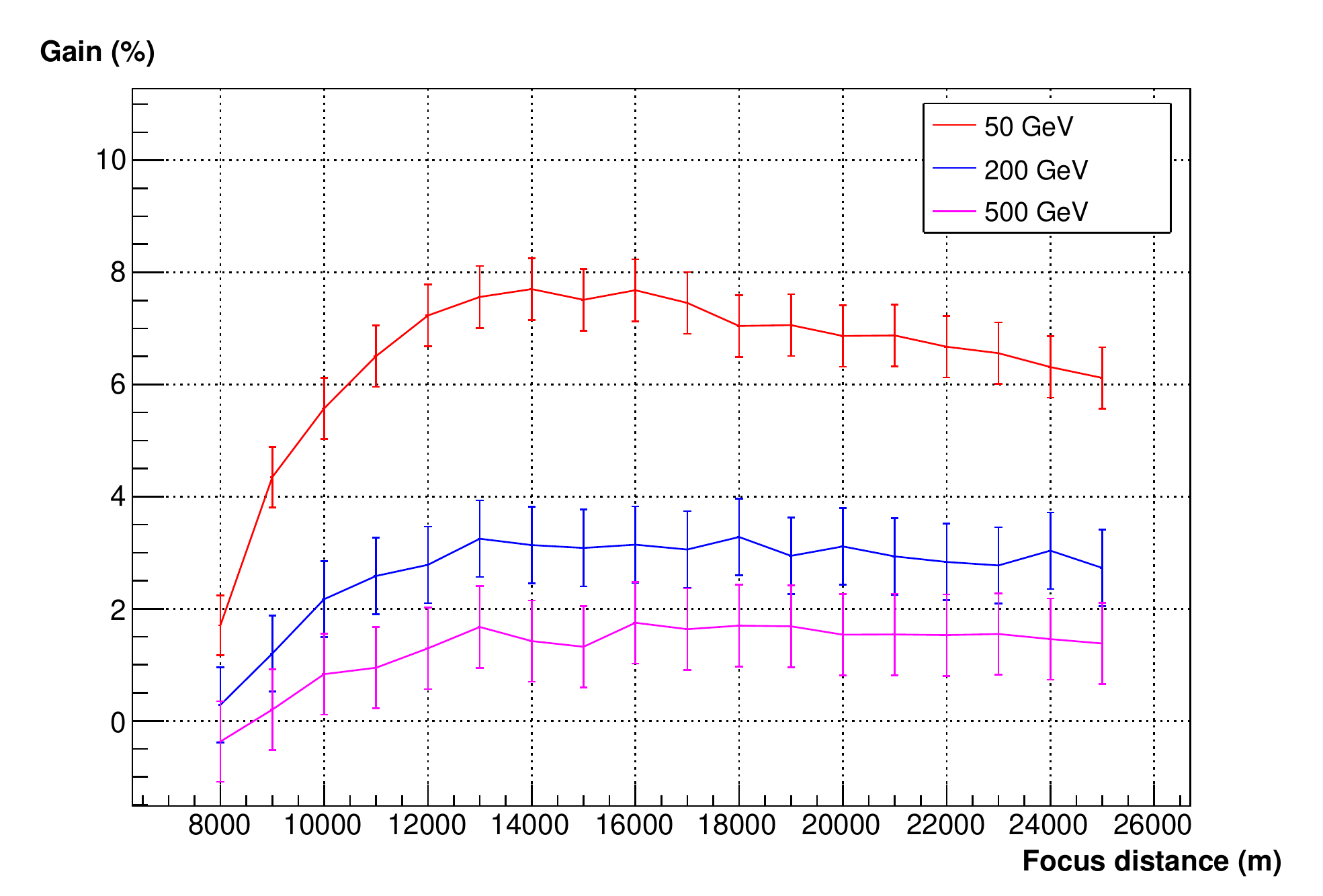}
  \caption{Trigger rate gain in percentage as a function of focus position and relative to the case of focusing to infinity. Gamma-ray samples of three different energies and for incident direction with zenith angle = 32$^{\circ}$.}
  \label{gain32}
 \end{figure}

\begin{figure}[h]
  \centering
  \includegraphics[width=0.5\textwidth, angle = 0]{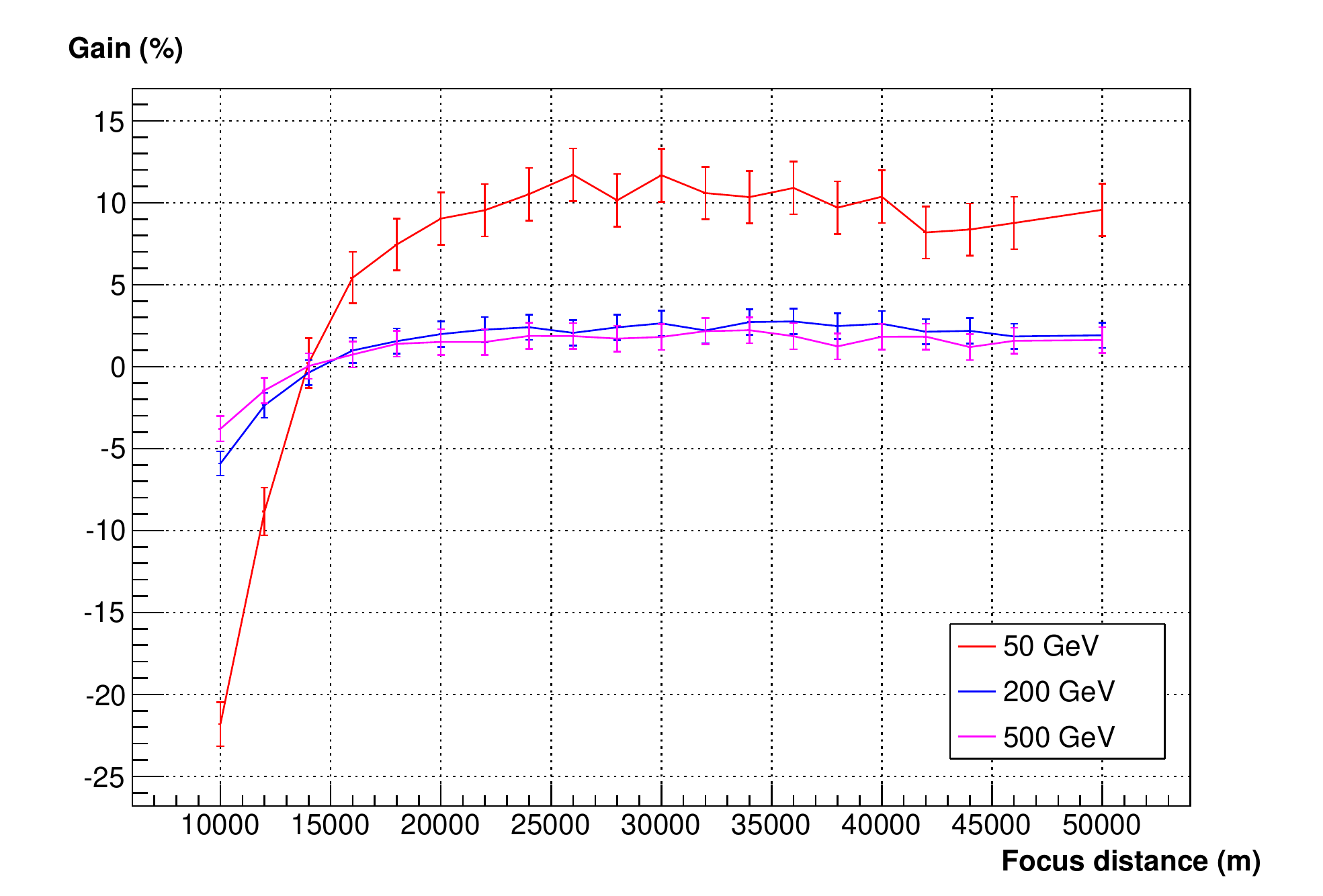}
  \caption{Trigger rate gain in percentage as a function of focus position and relative to the case of focusing to infinity. Gamma-ray samples of three different energies and for incident direction with zenith angle = 60$^{\circ}$.}
  \label{gain60}
 \end{figure}

Similar conclusions are achieved for other zenith angles (Figures \ref{gain32} and \ref{gain60}).
Besides there is an evolution of the best focus position from 12 km at ZA = 0$^{\circ}$ to 20-40 km at ZA = 60$^{\circ}$. 
On the whole, these results support the possibility to chose a relevant focus position for each zenith angle for a given observation. 

A second step consisted in introducing the reconstruction of the shower events to estimate the focus impact on the identification of the gamma-ray direction. For this study the geometrical Hillas reconstruction method \cite{bib:hillas} was used and applied only in mono-telescope mode.

\begin{figure}[h]
  \centering
  \includegraphics[width=0.5\textwidth, angle = 0]{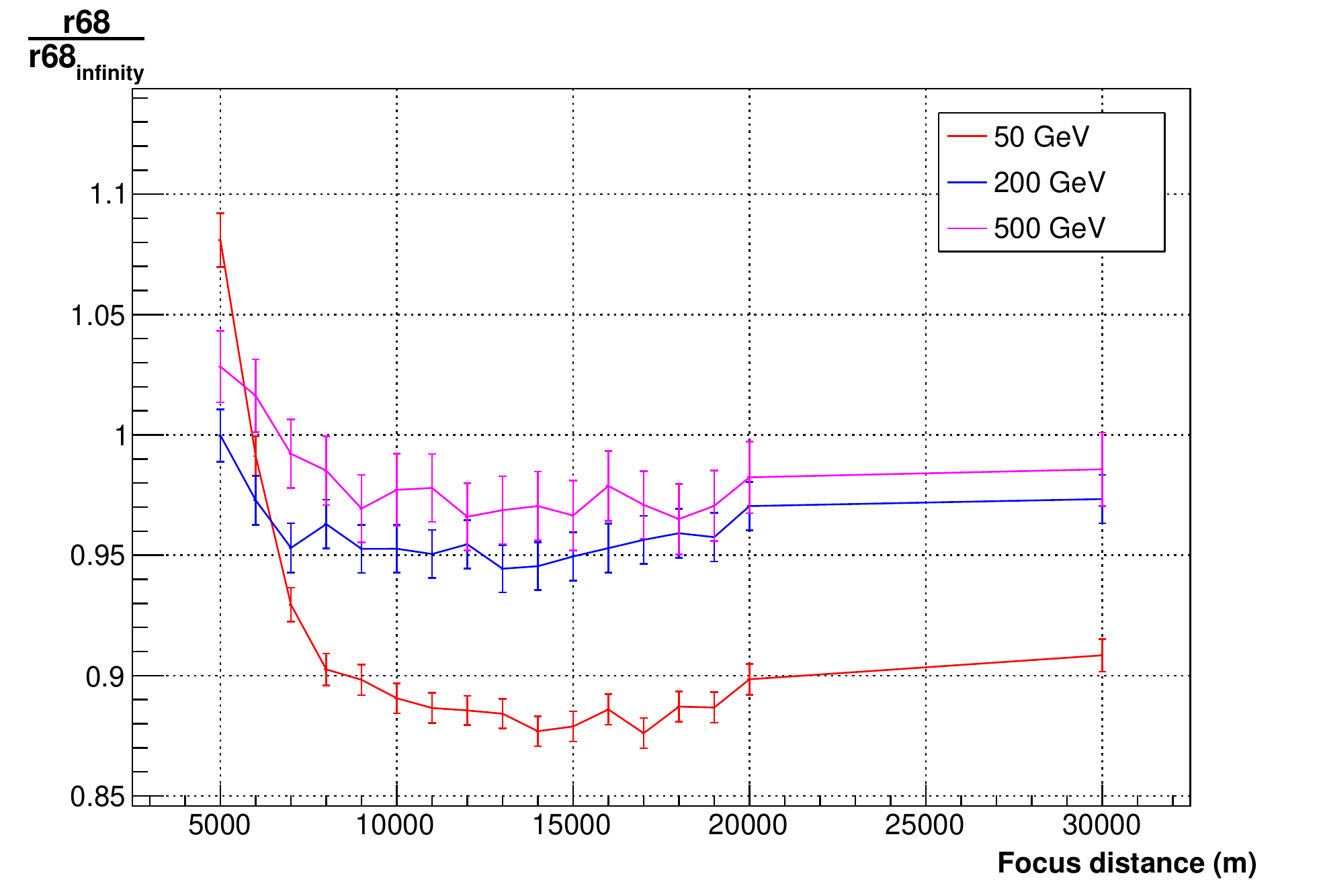}
  \caption{Relative improvement of the angular resolution as a function of focus distance with respect to the focus at infinity. Results correspond to zenith angle = 0$^{\circ}$ and three fixed energy gamma-ray samples (50 GeV, 200 GeV and 500 GeV).}
  \label{AngRes}
 \end{figure}

Figure \ref{AngRes} shows our results on the angular resolution for ZA = 0$^{\circ}$ .
The 68\% containment (r68) of the distribution of the difference between reconstructed and true direction of primary gamma rays generated from a point-like source is assumed to estimate our angular resolution. Choosing the best focus position implies  an improvement of the resulting r68, especially at low energy.
Again this improvement is very stable with the focus distance. Moreover, the results are identical for ZA = 32$^{\circ}$ and ZA = 60$^{\circ}$.


\section{Conclusion}

The effect of the focus position on the CT5 telescope performance as a function of energy and incident direction of the gamma rays has been studied. The results of the simulation studies confirm the critical role of the focus position on the trigger sensitivity and the angular resolution.
Furthermore the stability of the performance guarantees the possibility to choose the most appropriate focus position as a function of the observation configurations.

The best height range is between 12 km and 40 km corresponding to a focus length between 36.11 m and 36.03 m respectively. CT5 has the technical device allowing to fulfill this requirement.       

The above results indicate some margin of improvement of the gamma-to-hadron separation. This hypothesis as well as the validation of the MC simulation results are currently studied by using the first CT5 data.

\vspace*{0.5cm}
\footnotesize{{\bf Acknowledgment: }{please see standard acknowledgement in H.E.S.S. papers, not reproduced here due to lack of space.}}

\end{document}